\documentclass[%
 reprint,
 amsmath,amssymb,
 aps,
]{revtex4-2}
\usepackage{graphicx}% Include figure files
\usepackage{float} %floating figures
\usepackage{hyperref}
\usepackage{natbib}
\usepackage{xcolor} %change the color of text
\usepackage{ulem}
\usepackage{dcolumn}% Align table columns on decimal point
\usepackage{bm}% bold math
\begin{document}

\preprint{APS/123-QED}

\title{Correlated Insulating States in Twisted Double Bilayer Graphene Enhanced by Interfacial Effect on CrOCl}% Force line breaks with \\

\author{Ning Ma$^{1}$}
\author{Zekang Zhou$^{1}$}
\author{Chiara Cocchi$^{2}$}
\author{Maurice Bal$^{2}$}
\author{Maarten van Delft$^{2}$}
\author{Kenji Watanabe$^{3}$}
\author{Takashi Taniguchi$^{4}$}
\author{Steffen Wiedmann$^{2}$}
\author{Jian-Hao Chen$^{5}$}
\author{Mitali Banerjee$^{1,6}$}
\email{mitali.banerjee@epfl.ch}

\affiliation{$^1$ Institute of Physics, Ecole Polytechnique Fédérale de Lausanne (EPFL), CH-1015 Lausanne, Switzerland}

\affiliation{
$^2$ High Field Magnet Laboratory (HFML-EMFL), Radboud University, Toernooiveld 7, 6525 ED Nijmegen, The Netherlands}

\affiliation{
$^3$ Research Center for Functional Materials, National Institute for Materials Science, 1-1 Namiki, Tsukuba 305-0044, Japan}

\affiliation{
$^4$ International Center for Materials Nanoarchitectonics, National Institute for Materials Science, 1-1 Namiki, Tsukuba 305-0044, Japan}

\affiliation{
$^5$ International Center for Quantum Materials, School of Physics, Peking University, Beijing 100871, China}

\affiliation{
$^6$ Center for Quantum Science and Engineering (QSE Center), École Polytechnique Fédérale de Lausanne (EPFL), 1015, Lausanne, Switzerland}

\date{\today}% It is always \today, today,
             %  but any date may be explicitly specified

\begin{abstract}
\noindent 
Interaction between different two dimensional materials can give rise to many exotic physical phenomena which are rarely observed in intrinsic materials. Recently, several theoretical and experimental works have revealed that magnetic proximity effect between pristine graphene and magnetic substrates can lead to the emergence of quantum anomalous Hall states and quantum spin Hall states. However, interplay between correlated states in graphene-based systems and magnetic materials has seldom been studied. Here we perform the transport measurement at ultrahigh magnetic field of twisted double bilayer graphene (TDBG) on CrOCl (COC) substrate, which is an antiferromagnetic material. Instead of a magnetic-exchange effect on graphene, we observe an enhanced correlated insulating state at half-filling factor of TDBG as a result of the charge-transfer process between TDBG and COC. The temperature and magnetic field dependence of this enhanced state are further studied. Our results demonstrate the influence of charge-related effect at the interface, and shed a light on a new route for manipulating the correlated states in graphene-based moiré systems using interfacial engineering.
\end{abstract}

%\keywords{Suggested keywords}%Use showkeys class option if keyword
                              %display desired
\maketitle

%\tableofcontents

\noindent 
\centerline{\textbf{I. Introduction}}

Heterostructures assembled by different types of van der Waals materials have shown potential in investigation of exotic properties that cannot exist in single crystals. For example, when placing graphene in close proximity to magnetic substrates, it is predicted that numerous novel phases, i.e., quantum anomalous Hall phase\cite{qiao_quantum_2010,qiao_quantum_2014,zhang_robust_2015,hogl_quantum_2020} and quantum spin Hall phase\cite{kane_quantum_2005,kaloni_quantum_2014,hatsuda_evidence_2018,ghiasi_quantum_2024} can emerge in such systems, owing to the magnetic exchange or strong spin-orbital coupling in the substrates. However, since a series of experimental efforts have recently been made on graphene and antiferromagnetic materials, i.e.,graphene/CrX$_3$(X = Cl, Br, I)\cite{tang_magnetic_2020,tseng_gate-tunable_2022}, graphene/CrPS$_4$\cite{ghiasi_quantum_2024}, graphene/CrOCl\cite{wang_quantum_2022,yang_unconventional_2023}, graphene/CrSBr\cite{yang_electrostatically_2024} and so on\cite{wu_large_2020,yi_exploring_2023}, interfacial charge-related effects are found to play a crucial role in the interaction between graphene and magnets, which can dominate the proximity effects at the interface, tending to modulate the quantum Hall states in graphene by the proximitized charge transfer process. Theoretical works\cite{lu_synergistic_2023} have shown that charge transfer effect can yield a long-wavelength electronic crystal at the interface, which serves as a superlattice Coulomb potential on graphene, and further leads to a spontaneous bandgap opening in graphene\cite{yang_unconventional_2023,lu_synergistic_2023}.

The charge transfer process exists widely in those aforementioned graphene/insulator heterostructures, a question about the influence of the charges accumulated in the substrates on correlated states in graphene systems (i.e., twisted graphene family\cite{cao_correlated_2018,serlin_intrinsic_2020,polshyn_electrical_2020,xu_tunable_2021,li_imaging_2022,tseng_anomalous_2022,zhang_visualizing_2021,burg_correlated_2019,cao_tunable_2020,saito_independent_2020}) arises. In many different graphene-based moiré systems, TDBG is a promising platform\cite{burg_correlated_2019,cao_tunable_2020,liu_tunable_2020,he_symmetry_2021,liu_isospin_2022,wang_emergent_2024} to address this question, since its band structure is strongly tuned by the displacement field\cite{koshino_band_2019,chebrolu_flat_2019}. Coulomb interaction in TDBG can be highly modulated, which is also related to the topological properties in TDBG. The advantage of tuning the Chern numbers of those topological flat bands in TDBG, provides the possiblity to investigate the connection between correlated states in TDBG and the topology behind them.

In this work, we study the electrical transport of TDBG with a twist angle of around 1.1 degrees in proximity to the antiferromagnetic insulator CrOCl\cite{zhang_magnetism_2019,gu_magnetic_2022,zhang_spin-lattice_2023}. The charge transfer effect is observed to influence not only the high resistance state at the charge neutral point (CNP), but also the correlated insulating states at half-filling factors. In addition, we investigate the interplay of the charge transfer effect and Hofstadter bands at ultra high magnetic fields, which shows an unexpected enhanced resistance occurring at half-filling of electron-doped regions.

\noindent 
\centerline{\textbf{II. Results}}

The TDBG-COC heterostructure (as shown in Figs.1(a) and (b)) is fabricated using the standard cut and stack technique\cite{saito_independent_2020}. A large flake of bilayer graphene is cut into two pieces using AFM tips, then picked up one by one and stacked with a certain twist angle. The TDBG device mentioned in the main text has a twist angle of around 1.1 degrees, which is extracted from the resistance minima in Brown-Zak oscillations\cite{barrier_long-range_2020,huber_band_2022} (as shown in the Supplementary Materials). Although it slightly deviates from the magic angle of AB-AB stacked TDBG previously reported\cite{burg_correlated_2019,liu_tunable_2020,cao_tunable_2020}, correlated states at half-filling factors are still observed, which will be discussed in the following paragraphs.
\begin{figure}
      \centering
      \includegraphics[width=1\linewidth]{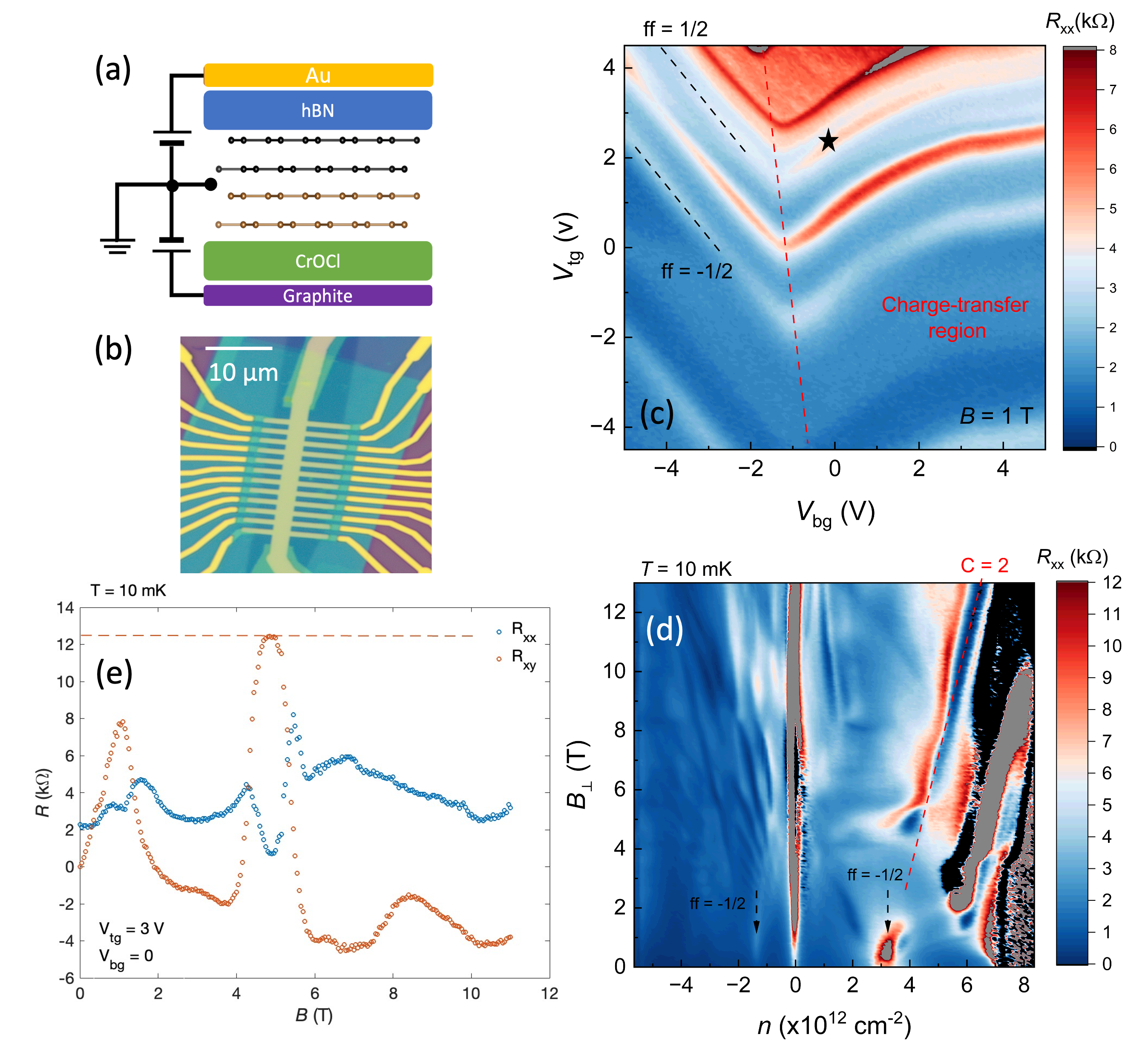}
      \caption{\label{fig:placeholder}Electrical measurement of TDBG-COC heterostructure under the charge-transfer effect. (a) Illustration of the structure of TDBG-COC sample. (b) Optical image of the device measured in the main text, which has been etched into Hall bar geometry by using O$_2$ and SF$_6$ plasma. The inset scale bar is 10 $\mu$m. (c) Longitudinal resistance as a function of dual gates measured at 10 mK and 1 T. The boundary of the region influenced by charge-transfer effect is indicated by red dashed line. Resistance peaks of filling factor(ff) = 1/2 and -1/2 are illustrated as the two black dashed lines. (d) Landau fan diagram measured at 10 mK, with D fixed at 0.26 V/nm. Half-filling states at zero field are indicated by the black arrows. The Chern insulator which stems from half-filling state is marked as the red dashed line.  (e) Longitudinal and Hall resistance of the Chern insulator versus magnetic field. These two lines are measured at the position marked by the black star in Fig. 1(c), which corresponds to 0.26V/nm of displacement field and 4x$10^{12}$cm$^{-2}$ of carrier density .}

\end{figure}

We performed the magneto-transport measurement in a dilution refridgerator at a base temperature of around 10 mK. Fig.1(c) shows the longitudinal resistance as a function of dual gates under 1 T perpendicular magnetic field. The dual gate map is separated into two different regions (indicated by the red dashed line in Fig.1(c)) by the prominent charge-transfer (CT) effect at the interface of TDBG and COC. On the left side of the map, the insulating states from Dirac point and half-filling are observed,as shown in Fig.1(c). On the right side of the charge-transfer boundary, all the resistance strips are bent, indicating that the electrons are injected into COC, which makes TDBG in the bottom right part of the map hole-doped. Then we fix the displacement field at 0.26 V/nm, plotting the Landau fan diagram measured from 0 to 13 T, as shown in Fig.1(d). Brown-Zak oscillations and Landau levels emanating from half-filling states and CNP are observed. On the electron side of the fan diagram, where the charge-transfer effect dominates, only a correlated insulating state from half-filling factor is found. A Chern insulator that stems from the half-filling state occurs from around 4 T and survives to a high field of 13 T. In order to calculate the Chern number of this Chern insulating state, both longitudinal and Hall resistance are measured as functions of magnetic field in Fig.1(e). C = 2 is extracted from the Hall plateau at 5 T, as shown in Fig.1(e), which agrees with previous literature of TDBG\cite{liu_isospin_2022}. 
\begin{figure}
    \centering
    \includegraphics[width=1\linewidth]{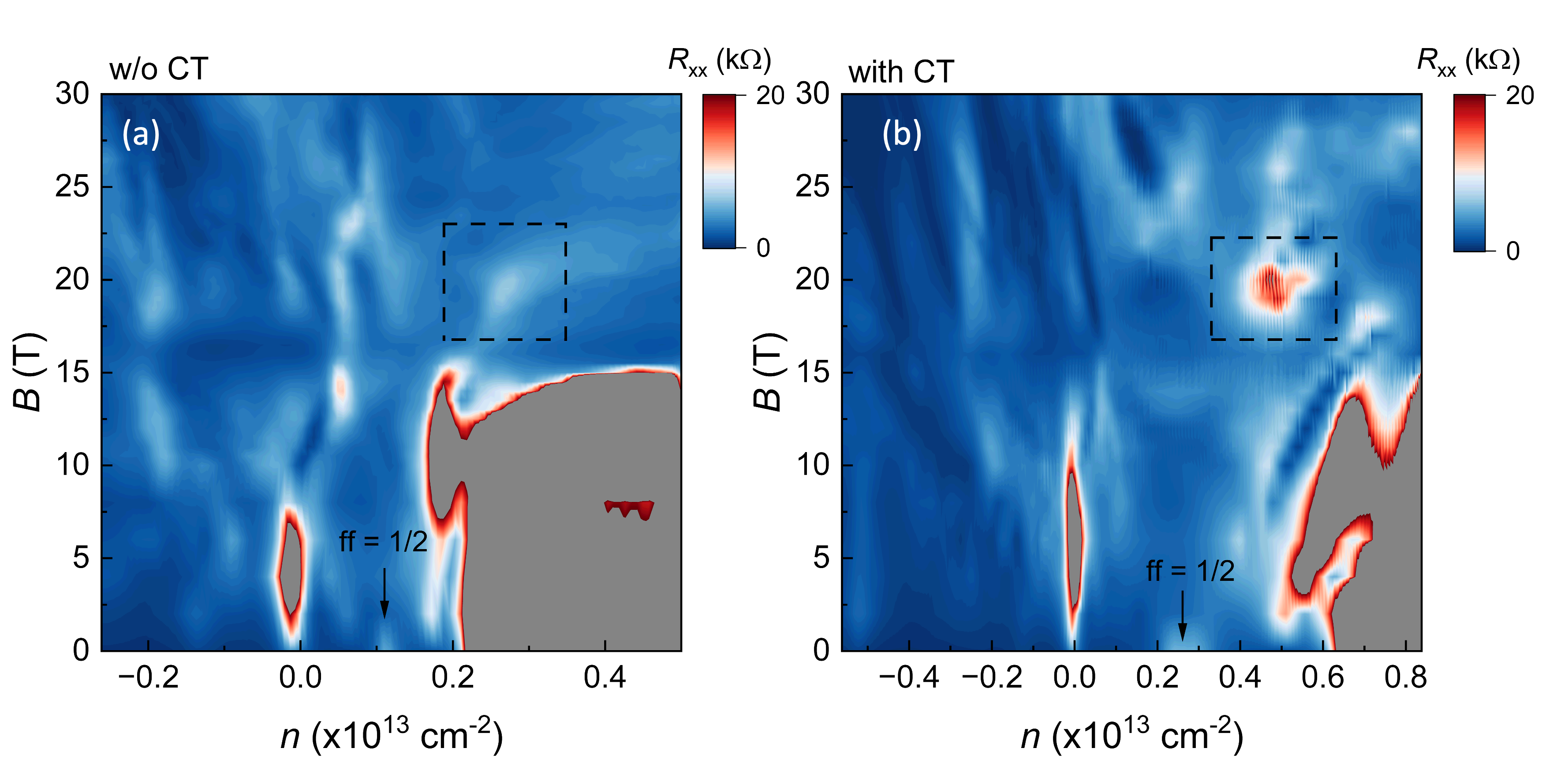}
    \caption{\label{fig:placeholder}Charge transfer-enhanced correlated states at half filling factor under high magnetic field. (a) and (b) The Landau fan diagrams of longitudinal resistance at D = 0.5 V/nm and 0.26 V/nm, corresponding to the regions without or with charge-transfer effect, respectively. The insulating states at ultrahigh magnetic field are shown in the black boxes. The insulating states in the "charge-transfer" region is enchanced, compared with the one in Fig. 2(a).}
    
\end{figure}

So far, we have found that the main impact of the magnetic substrate on TDBG is about the charge-transfer effect, which influences the gate tunability of TDBG by injecting electrons into the substrate. However, no phenomenon related to the magnetic properties of COC has been observed in the device. In order to further investigate the effect of COC on TDBG, we plot and compare the complete Landau fan diagrams (as shown in Fig.2) of different displacement fields D = 0.5 V/nm and D = 0.26 V/nm, which correspond to the regions of ''without'' and ''with'' charge-transfer effect, respectively. By comparison, the features of both diagrams on the hole-doped side are quite similar. Hofstadter butterfly structures can be observed at high field. Detailed Landau fan diagrams of longitudinal and Hall resistance can be found in Fig.S4 in the Supplementary Materials. Landau levels reoccur at around 16 T, which corresponds to the case when half of the magnetic flux goes through one moiré unit cell\cite{das_observation_2022,herzog-arbeitman_reentrant_2022}. However, the behaviors on the electron side are found to be quite distinct.  The Chern insulators indicated by the red dashed line in Fig.1(d) are not clearly observed in the ''without charge-transfer'' region in Fig.2(a) at low magnetic field, although the half-filling insulating state at zero field can be observed in both diagrams. In addition, a strongly enhanced insulating state is found at around 20 T, as shown in the black box in Fig.2(b). According to the results in Fig.4(a), this insulating state is demonstrated as the half-filling state measured at zero field before, which reoccurs at ultrahigh field. Although the same insulating state can also be found in Fig.2 (a), the resistance in Fig.2(b) is remarkably enhanced as a result of the charge-transfer effect. One possible origin of the enhanced insulating state is the reconstruction of band structure induced by the e-e interaction between graphene and COC\cite{lu_synergistic_2023,yang_unconventional_2023,wang_quantum_2022}. The charge-transfer effect yields a long wavelength electronic crystal at the surface of COC, which provides an interlayer superlattice Coulomb potential to the graphene layers and triggers gap opening or being enhanced. Previous experiments have shown that in bilayer graphene-COC case, the interacting effect can be further enhanced by the interfacial coupling to the long-wavelength charge order at the surface of COC\cite{yang_unconventional_2023,lu_synergistic_2023}. On the other hand, it is also reported that strong magnetic proximity effect between graphene and magnetic substrates can induce gap opening by direct band structure hybridization\cite{cardoso_strong_2023,shi_magnetic_2023}, which is very sensitive to the magnetic configuration of the substrates and the relative orientation between the two layers. To find out the underlying mechanism behind the enhanced insulating state in our samples, further experiments are needed.

Furthermore, we measure the temperature and magnetic field dependence of the enhanced resistance of half-filling states versus varying carrier density, at fixed D = 0.26 V/nm . Fig.3(a) shows the case under perpendicular field of 20 T, the resistance of the half-filling state decreases with increasing temperature from 3.2 K to 20 K. The position of the resistance maxima also changes slightly at higher temperatures. By fitting the temperature dependence of the resistance maxima at different magnetic fields with Arrhenius formula, we calculate the thermal activation gap $\Delta$ under different magnetic fields, and estimate the g factor of the half-filling states, which is plotted in Fig.3(b). The gap $\Delta$ for both cases (with or without charge-transfer effect, as mentioned above), increases with the magnetic field and reaches its maximum at around 20 T. Then under a higher magnetic field, the gap gradually decreases, indicating a competition between valley and spin polarization. Thus, we chose the fitting windows for the valley-polarized gap within the range of magnetic field where the gap is developing and the valley polarization is dominant.  We compare g factors in the regions of “without charge-transfer” (g = 11.8) and “with charge-transfer” (g = 18.1), which also proves that the reentrant correlated insulating state is valley-polarized, consistent with the correlated state at low magnetic field. Additionally, the g factor of the latter (18.1) is much larger than the former one, which may indicate that the charge transfer effect has a strong impact on the enhancement of valley polarization of TDBG, together with the out-of-plane magnetic field.

\begin{figure}
    \centering
    \includegraphics[width=1\linewidth]{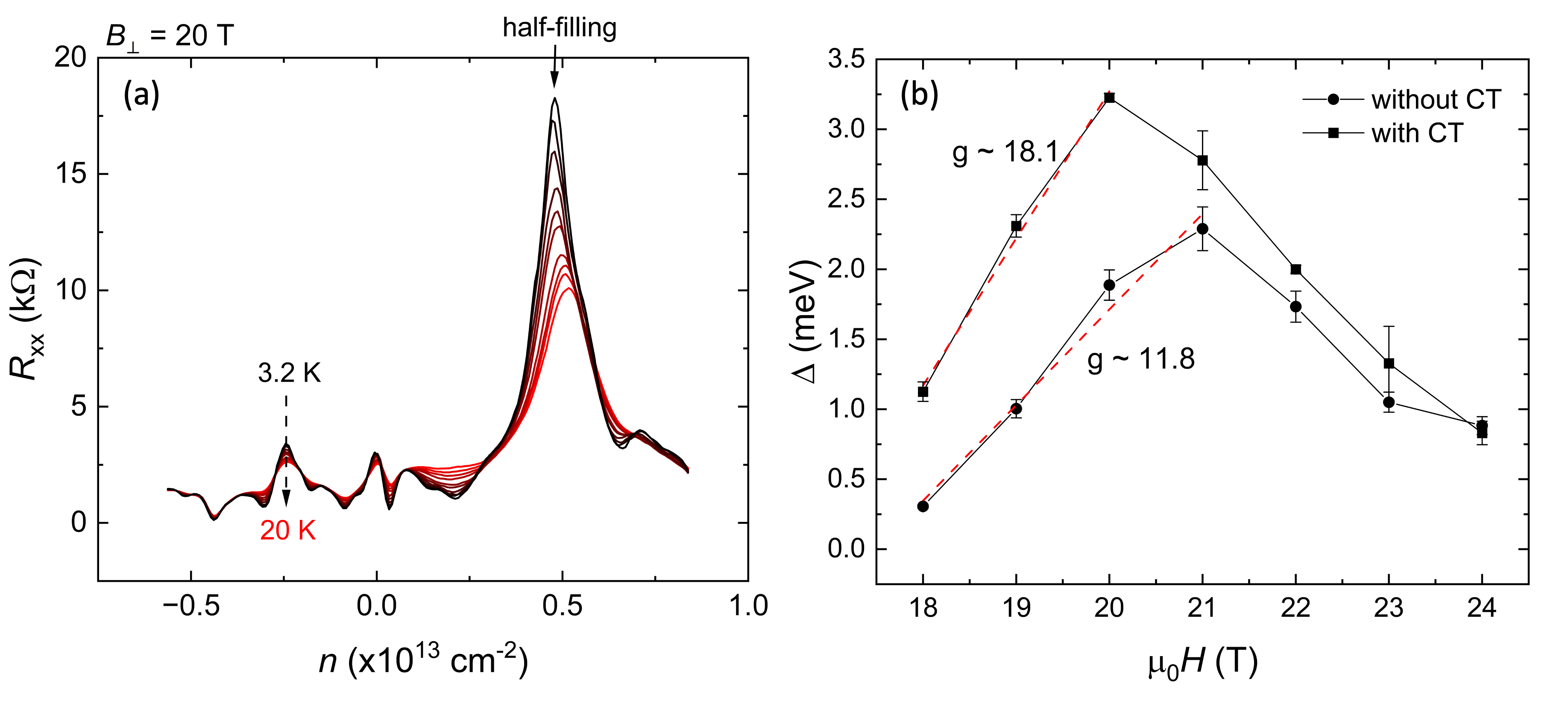}
    \caption{\label{fig:placeholder}Valley polarized correlated insulating state at half-filling factor under high magnetic field. (a) Temperature dependence of the correlated states at different carrier densities under 20 T. The resistance of the half-filling state with charge-transfer effect decreases with elevated temperature from 3.2 K to 20 K. (b) Estimation of the thermal activation gap of the insulating state at half-filling with or without the effect of charge transfer, respectively. The g factors of half-filling state with (18.1) or without (11.8) charge-transfer(CT) effect, respectively, are extracted from the slope of the relationship $\Delta$ versus out-of-plane magnetic field. }
\end{figure}

Besides, we study the behaviors of the correlated insulating states under in-plane field when the out-of-plane field fixed at 20 T. The results are shown in Fig.4. Apart from half-filling states, the insulating state at quarter-filling is also observed in Fig.4(a). When increasing in-plane magnetic field, the half-filling states are drastically suppressed, which indicates that the gap is gradually being closed under in-plane field. In contrast, the quarter-filling state survives even under 11 T. Usually the valley-polarized states are believed to be independent of in-plane field, our results show that there may exist interaction between the valley and spin degree of freedom. 

\begin{figure}
    \centering
    \includegraphics[width=1\linewidth]{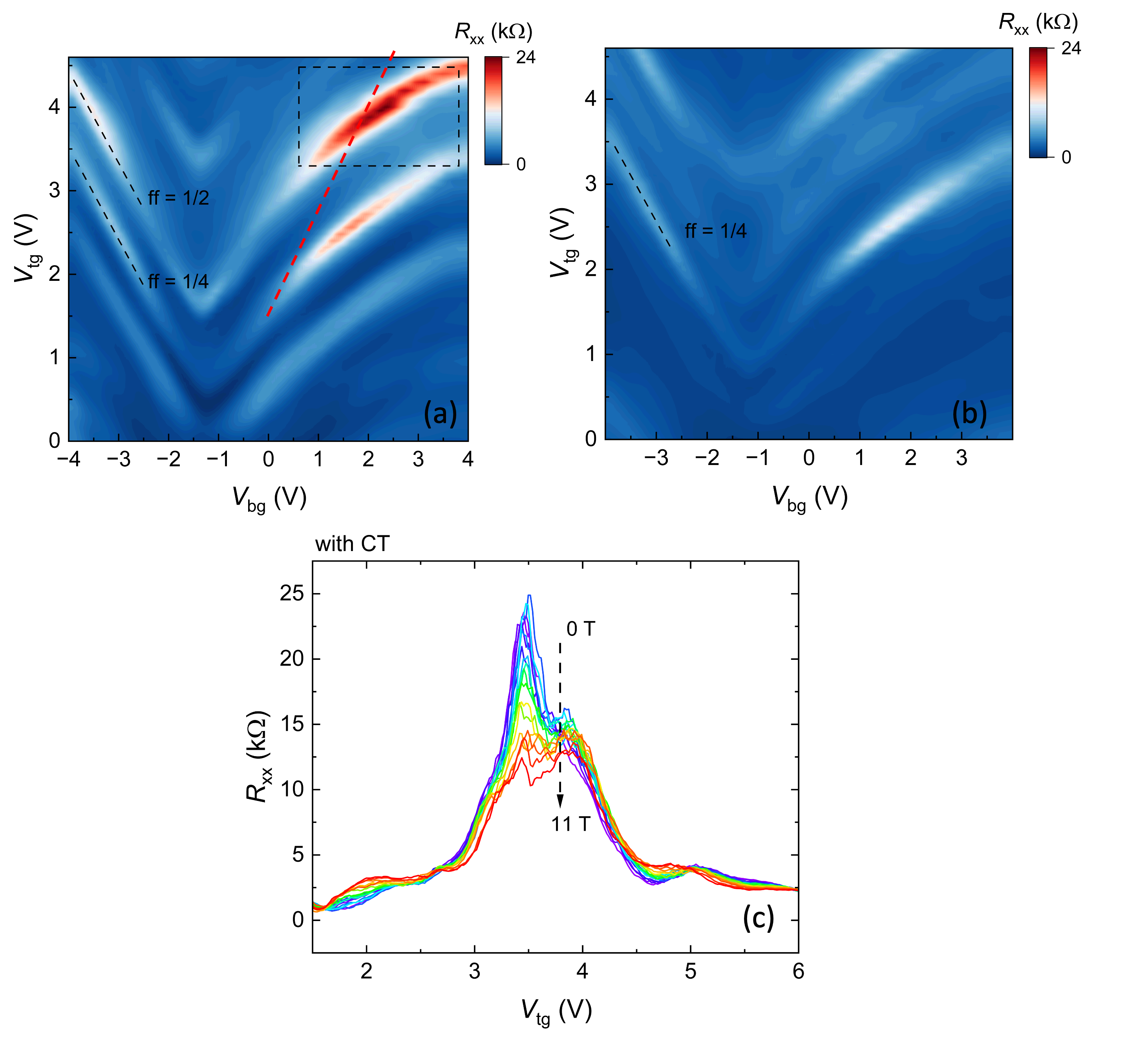}
    \caption{\label{fig:placeholder}In-plane field dependence of insulating states under high magnetic field. (a) Dual gate map measured at 4.2 K and 20 T of perpendicular field.  (b) Same dual gate map as (a) measured at 4.2 K, but under 20 T and 11 T of in-plane field. Half and quarter-filling states are indicated by the black dashed lines in (a) and (b).(c) The evolution of resistance of half-filling state in the black dahsed box in (a), with increasing in-plane field from 0 to 11 T. Gate voltages are scanning along the red dashed line indicated in fig.4 (a).}
   
\end{figure}
\vspace{1em}
\noindent
\centerline{\textbf{III. Conclusions}}

In summary, we have investigated the charge-transfer effect between TDBG and magnetic substrate. Reentrant correlated insulating states in TDBG are observed at ultrahigh magnetic field, similar to the case of twisted bilayer graphene. Furthermore, the valley-polarized half-filling states can be enhanced as a result of the charge-transfer effect. 
%Different responses of correlated states at either half-filling or quarter-filling to the in-plane field unveil the …. (mechanism?) 
For the first time, the influence of charge-transfer effect on the correlated states in moiré systems has been studied. Our results show the importance of charge-transfer effect in a hybrid system and may open a new pathway to manipulating the correlated states in moiré systems by interfacial engineering.

\vspace{1em}
\noindent
\textbf{Appendix A: Gate Sweeping Reproducibility of}
\\ \centerline{\textbf{the Enhanced Insulating State}}

It is known that different gate sweeping rates, temperature variation, and even measurement history may have impact on the interfacial charge-transfer effect. In order to prove the reproducibility of the appearance of the enhanced insulating state, we performed the measurements with different sweeping rates and also provide the results measured in a different coolingdown cycle after four months, giving evidence of the robustness of the insulating state and excluding the influence of the gate-sweeping related issues.

As shown in Fig.5, the zoom-in Landau fan diagram was measured at 0.6 K, with a smaller gate-sweeping step of 0.02 V. In comaparison, Fig.2(a) and (b) were measured with 0.05 V in each step. The enhanced insulating state appears in the same position in both Fig.5 and Fig.2(b), indicating that reproducibility of the insulating state under different scanning rates. Moreover, we repeated the measurement after four months since we found the main result in Fig.2. As shown in Fig.6, a Landau fan diagram is plotted at 4.2 K and with a gate-sweeping step of 0.04 V. The results are consistent with the ones in Fig.2(b) and Fig.5,  giving the enhanced insulating state in the same carrier density and magnetic field. The resistance value of the insulating state in Fig.6 is also in good agreement with the temperature-dependent data in Fig.3(a).

For the temperature variation, one can find in Fig.3(a) that resistance peak at 20 T survives from 3.2 K to 20 K, with the corresponding carrier density unchanged, which rules out the influence of temperature-induced charge-transfer effects.

\begin{figure}
    \centering
    \includegraphics[width=0.8\linewidth]{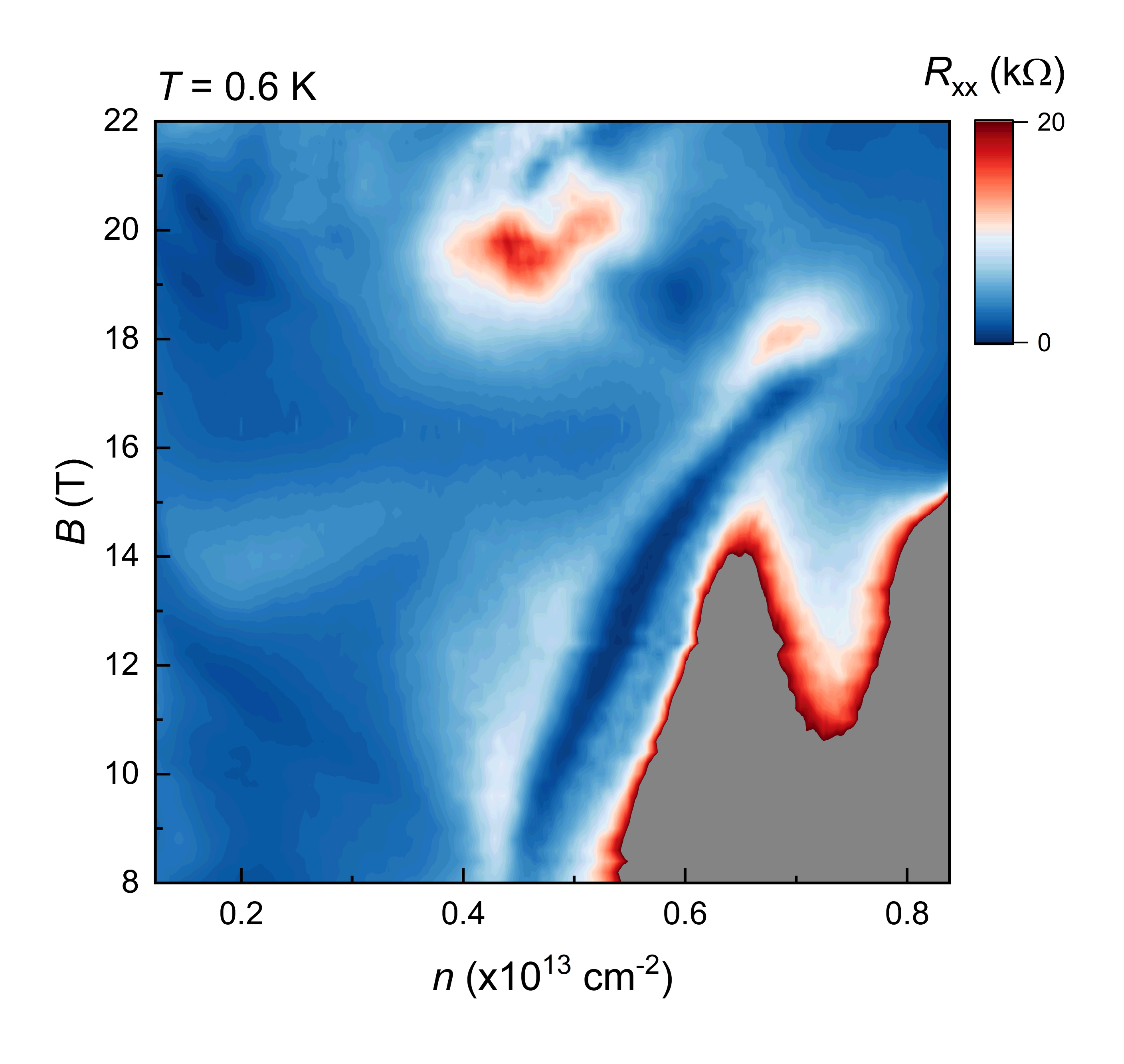}
    \caption{\label{fig:placeholder} Zoom-in Landau fan diagram measured at 0.6 K, with a gate-sweeping step of 0.02 V. Compared with the features in Fig.2(b), the enchanced insulating state occurred at the same carrier density and magnetic field. Besides, the resistance of the enhanced insulating state in both figures are comparable. It proves the enhanced insulating state is reproducible under different scanning rates of gate. }
   
\end{figure}

\begin{figure}
    \centering
    \includegraphics[width=0.8\linewidth]{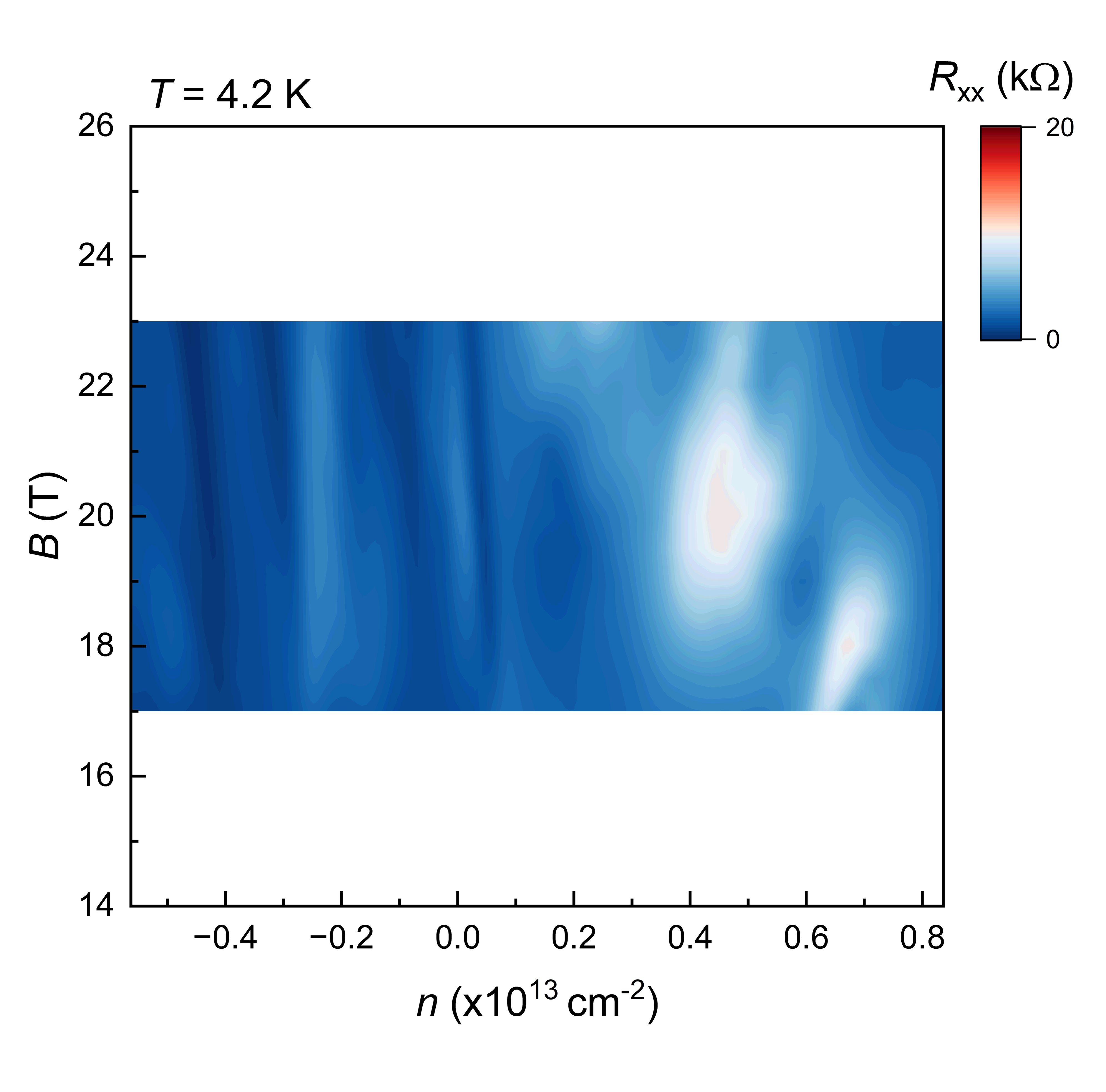}
    \caption{\label{fig:placeholder} Zoom-in Landau fan diagram measured at 4.2 K, with a gate-sweeping step of 0.04 V. This Landau fan diagram was measured after four months since we observed first time the enhanced insulating states, to prove the robustness of the device. The insulating state is visible in the same position compared with Fig.2(b) and Fig.5. The resistance at 20 T is of the same value as the one given in Fig.3(a), which indicates the reproducibility.}
   
\end{figure}

\vspace{1em}
\noindent
\textbf{Appendix B: Definition of $n$ and $D$ in the region}
\\ \centerline{\textbf{with charge-transfer effect}}

Since it is hard to measure the contribution of carrier density in the substrate experimentally, the $n$ and $D$ in charge-transfer region are the total carrier density and effective displacement field considering contributions from both graphene and CrOCl. The $n$ and $D$ are defined by $n$=$(C_{tg} V_{tg}+C_{bg} V_{bg})/e-n_0$ and $D$=$(C_{tg} V_{tg}-C_{bg} V_{bg})/(2\epsilon_0 )-D_0$, respectively. Here $C_{tg}$ and $C_{bg}$ are the capacitances per area of top and bottom gate, respectively. $n_0$ and $D_0$ are the offsets from the residual carrier density and displacement field, respectively.

%\noindent \textbf{Data availability}

%\noindent

\

%\noindent \textbf{Code availability}

%\noindent Source code used to perform the calculations in this paper is available from the corresponding author upon request.

%\nocite{*}

\newpage

\bibliography{references}% Produces the bibliography via BibTeX.

%\noindent 

\noindent \textbf{Acknowledgements}

We thank J.P.Liu for helpful discussions. We acknowledge the support of the HFML, member of the European Magnetic Field Laboratory (EMFL). M.B. acknowledges the support of SNSF (Eccellenza Grant No. PCEGP2\_194528) and support from the QuantERA II Programme, which has received funding from the European Union’s Horizon 2020 research and innovation programme (Grant Agreement No. 101017733). K.W. and T.T. acknowledge support from the JSPS (KAKENHI Grant Nos. 20H00354 and 23H02052) and World Premier International Research Center Initiative, MEXT, Japan. G.W. is supported by the SNSF (Ambizione Grant No. PZ00P2-216183). 

\noindent \textbf{Author Contribution}

M.B. and N.M. conceived the project. N.M. made the stacks and fabricated the devices with the help of Z.Z. N.M. performed the measurements with the help of Z.Z. N.M. performed the high magnetic field measurements with the help of S.W., C.C., M.B. and M.v.D. N.M. has analyzed the data with inputs from M.B. and J.-H.C. K.W. and T.T. provided the hBN crystals. J.-H.C. provided the COC crystals. N.M. wrote the manuscript with inputs from M.B., J.-H.C., as well as all other authors.

\noindent \textbf{Competing Interests}

\noindent The authors declare that they have no competing interests.

\newpage

%\noindent \textbf{Figure captions}

%\label{fig1}

\end{document}